\documentclass[11pt,twoside]{article}
\usepackage{./asp2014}

\aspSuppressVolSlug
\resetcounters

\begin{document}

\title{On the Interaction in a Quartet of Galaxies}
\author{A.~A.~Yeghiazaryan, A.~A.~Hakobyan, and T.~A.~Nazaryan
\affil{Byurakan Astrophysical Observatory, 0213 Byurakan, Aragatsotn province, Armenia; \email{hakobyan@bao.sci.am}}}

% This section is for ADS Processing.  There must be one line per author.
\paperauthor{A.~A.~Yeghiazaryan}{anahit_y@hotmail.com}{}{Byurakan Astrophysical Observatory}{}{Byurakan}{Aragatsotn province}{0213}{Armenia}
\paperauthor{A.~A.~Hakobyan}{hakobyan@bao.sci.am}{0000-0001-7392-1765}{Byurakan Astrophysical Observatory}{}{Byurakan}{Aragatsotn province}{0213}{Armenia}
\paperauthor{T.~A.~Nazaryan}{nazaryan@bao.sci.am}{}{Byurakan Astrophysical Observatory}{}{Byurakan}{Aragatsotn province}{0213}{Armenia}

\begin{abstract}
We performed the Fabry-Perot
scanning interferometry of the quartet of galaxies
NGC~7769, 7770, 7771 and 7771A in H$\alpha$ line
and studied their velocity fields.
We found that the rotation curve of NGC~7769 is weakly distorted.
The rotation curve of NGC~7771 is strongly
distorted with the tidal arms caused by
direct flyby of NGC~7769 and flyby of a smaller neighbor NGC~7770.
The rotation curve of NGC~7770 is significantly skewed because of
the interaction with much massive NGC~7771.
The rotation curves and morphological disturbances
suggest that the NGC~7769 and NGC~7771
have passed the first pericenter stage,
however, probably the second encounter has not happened yet.
\end{abstract}

\section{Velocity fields and rotation curves}

We report the results of the optical interferometry of the interacting system
of galaxies NGC~7769, 7770, 7771 and 7771A and analyze their kinematics.
A detailed description of the morphological
features of the galaxies as well as photometry and color analysis of NGC~7769
are presented in the complete version of the study:
\citet{2015arXiv151000193Y}.
We also discuss the influence of interaction on
the kinematics, dynamics and star formation in the system.
Known models of galaxy interactions are based mostly on statistical observational data.
We try to illustrate how and to what extend these models can be applied to
explain the features of the galaxies in this system.

In order to study the velocity fields of the galaxies,
the observations were carried out at the 2.6m telescope of
the Byurakan Astrophysical Observatory (BAO, Armenia)
on 8 November 1996, with the ByuFOSC
(Byurakan Faint Object Spectral Camera) in the interferometric
mode, attached at the prime focus of the telescope.

\begin{figure}[hb]
\begin{center}$
\begin{array}{@{\hspace{0mm}}c@{\hspace{1mm}}c@{\hspace{0mm}}}
\includegraphics[width=0.47\hsize]{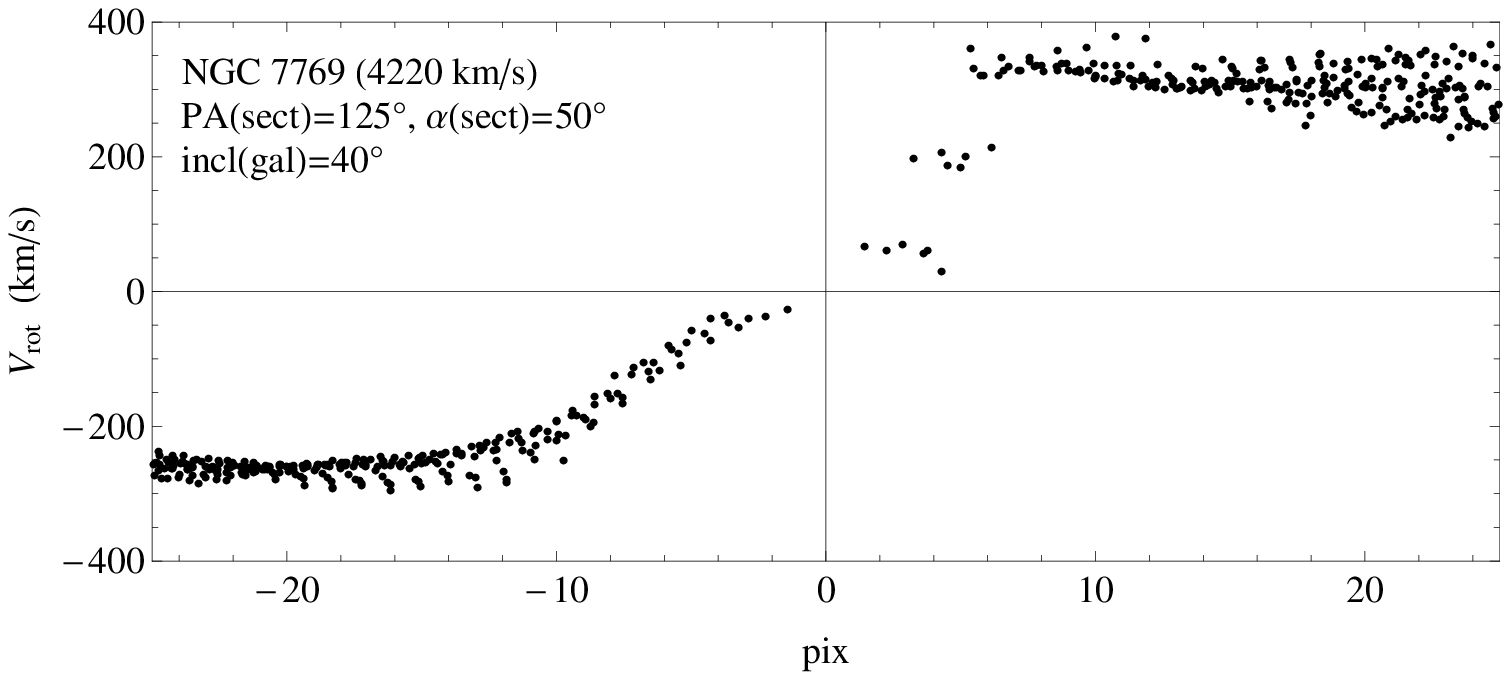} &
\includegraphics[width=0.52\hsize]{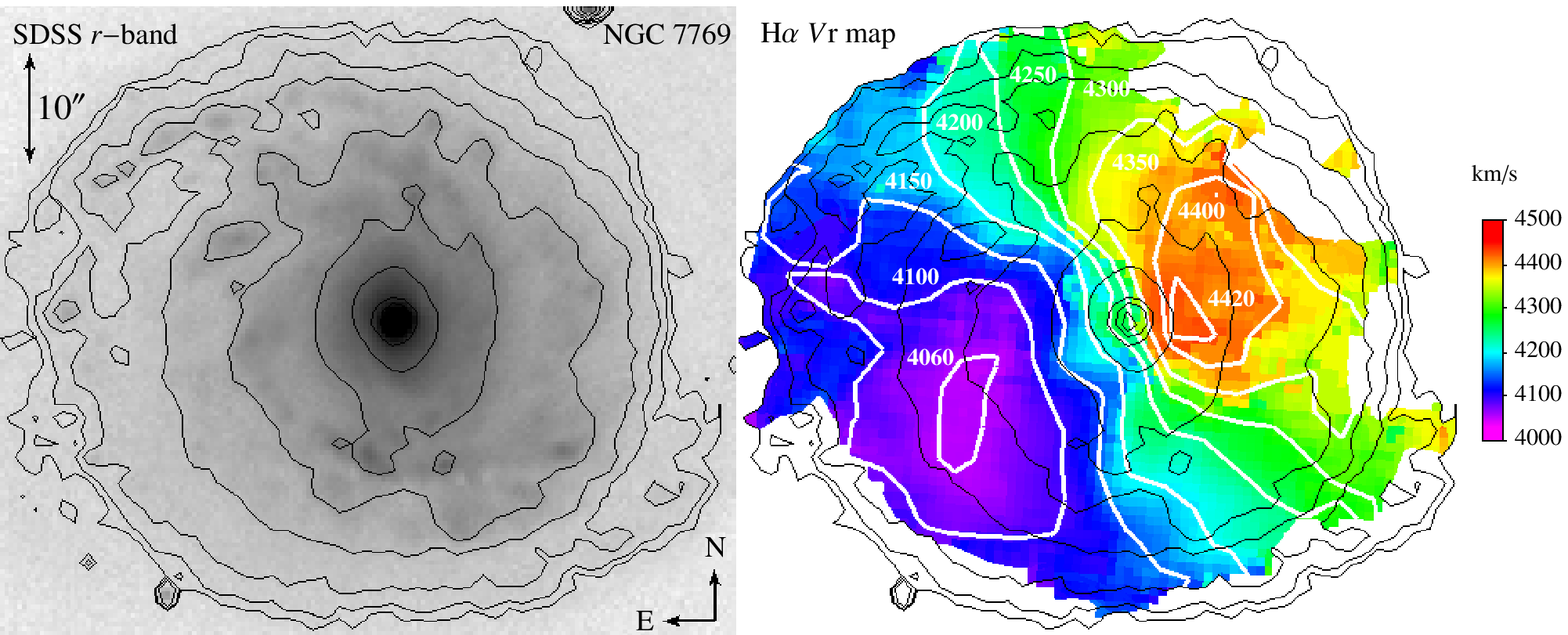}\\
\includegraphics[width=0.47\hsize]{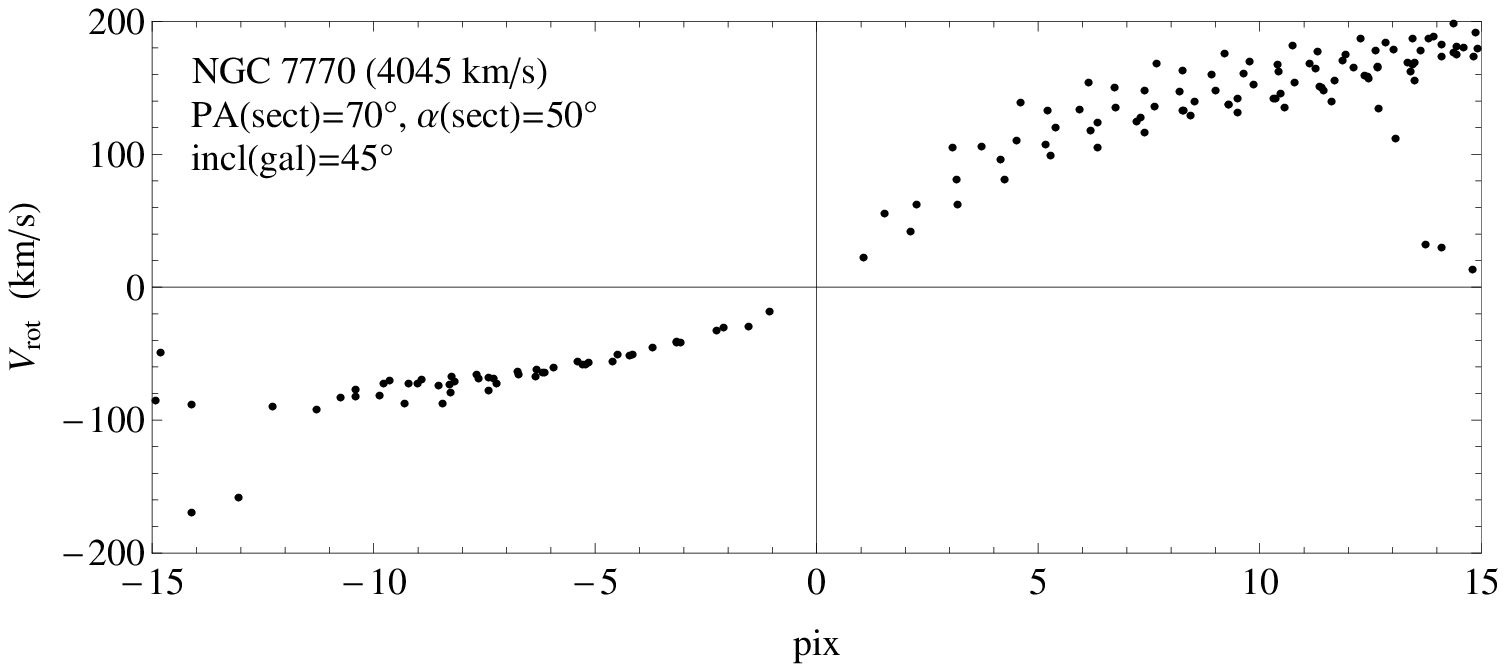} &
\includegraphics[width=0.52\hsize]{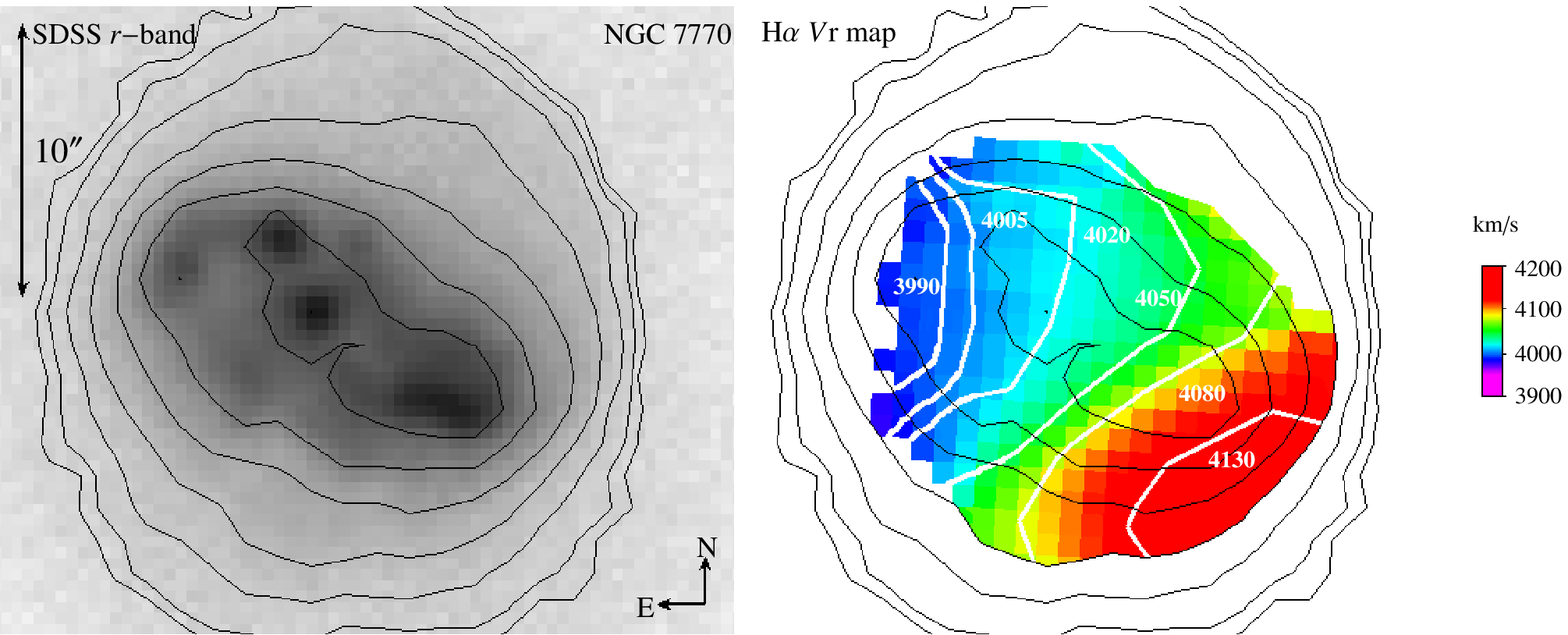}\\
\includegraphics[width=0.47\hsize]{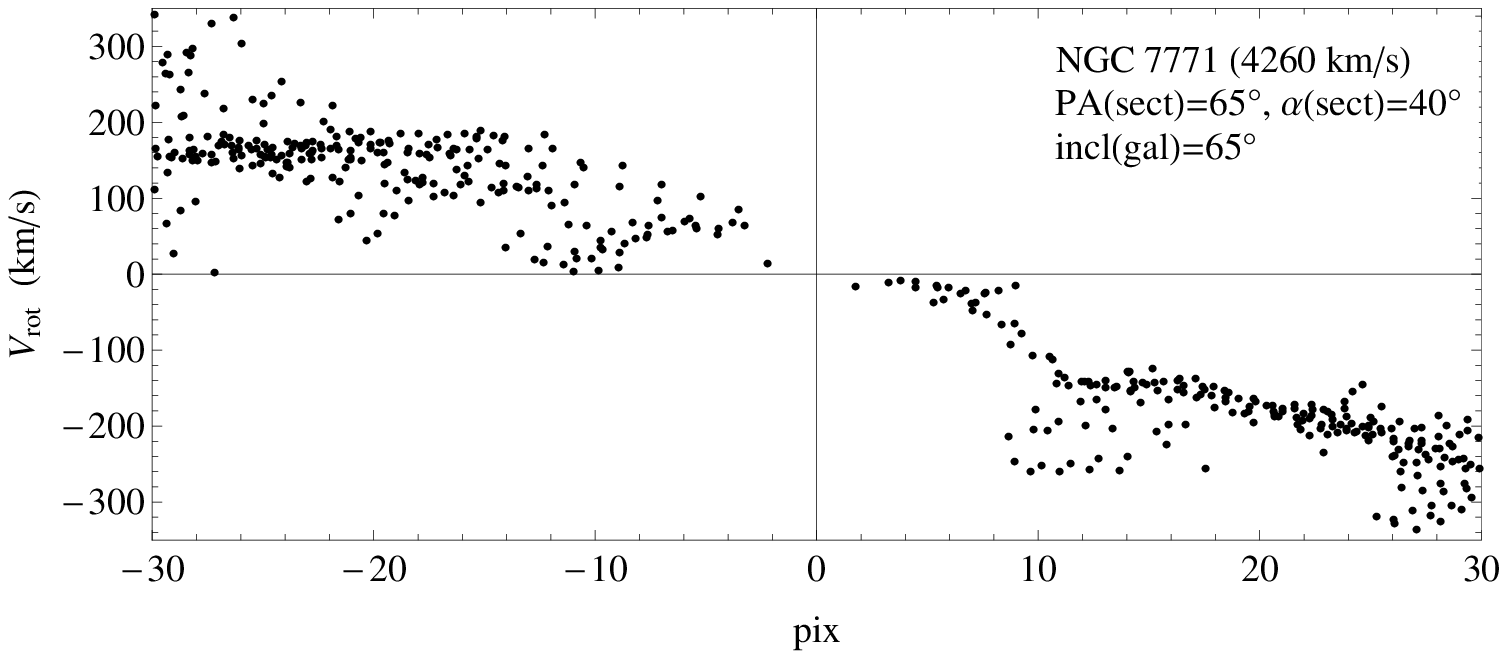} &
\includegraphics[width=0.52\hsize]{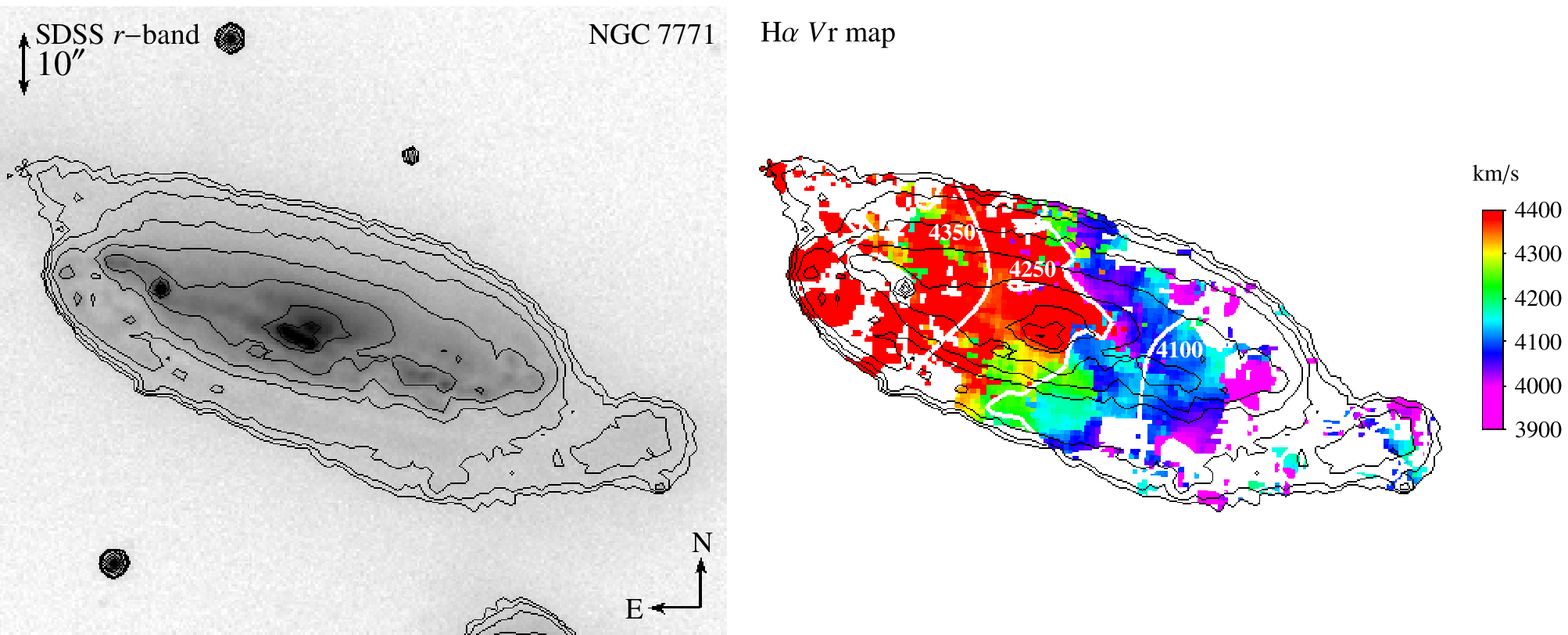}\\
\includegraphics[width=0.47\hsize]{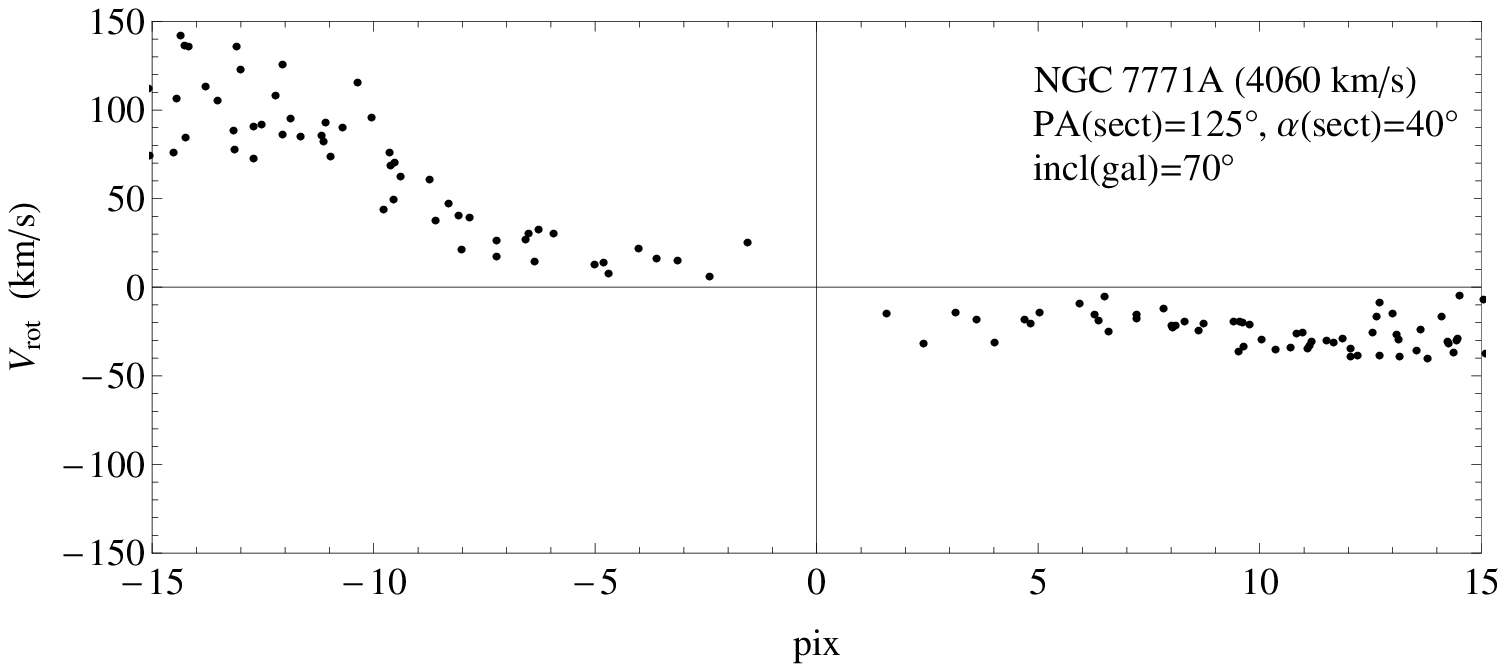} &
\includegraphics[width=0.52\hsize]{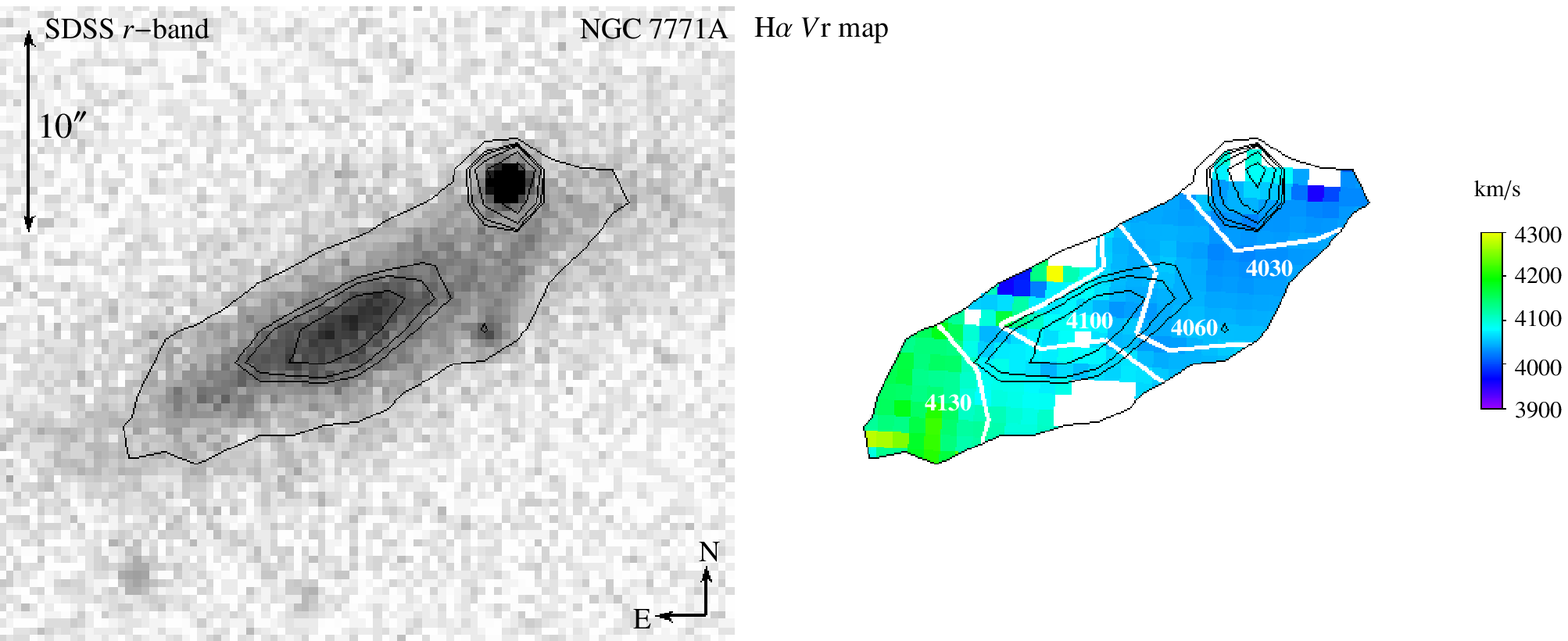}
\end{array}$
\end{center}
\caption{\emph{Right:} H$\alpha$ velocity fields of galaxies NGC~7769, 7770, 7771 and 7771A,
overlapped by the SDSS \emph{r}-band isophotes (black), and isovelocity contours (white).
The outer isophote corresponds to 22 ${\rm mag~arcsec^{-2}}$ for NGC~7769, 7770 and 7771,
and to 23 ${\rm mag~arcsec^{-2}}$ for NGC~7771A.
\emph{Left:} Derived rotational curves of the galaxies. One pixel on the horizontal axis
corresponds to 0.77 arcsec. For each plot, the radial velocity of the galaxy center,
the angle and PA of the sector used to obtain velocity data,
as well as the inclination of galaxy used in the calculations are shown.}
\label{vrot_velr}
\end{figure}

Based on the H$\alpha$ velocity fields (the right-hand panels of Figure~\ref{vrot_velr}),
we calculated the rotation curves of the galaxies (the left-hand panels of Figure~\ref{vrot_velr})
by using data points within sectors along the maximal gradient direction,
see isovelocity contours in the right-hand panels of Figure~\ref{vrot_velr}.

Maximal rotational velocity of NGC~7769 is observed at the radius of
around 15~arcsec from the galaxy nucleus.
The rotational velocities in Figure~\ref{vrot_velr} are in good agreement
with the HI measurements ($316~{\rm km \, s^{-1}}$) in \citet{1993ApJ...419...30C}.
Our measurements of velocities, having a better spatial resolution
compared with those of the previous studies (\citealt{1993ApJ...419...30C,1997AJ....114...77N}),
reveal weak perturbations of the rotation curve of NGC~7769,
which may be caused by interaction with NGC~7771.

The same cannot be said about the velocity field of NGC~7771.
Figure~\ref{vrot_velr} shows that there are perturbations and
large dispersion in radial velocities at the distances
larger than about 10-15 arcsec from the nuclei.
This distance is about half radius of the bar.
Evidently, this scatter of radial velocities can be explained
by the fact that part of the arms are included in the sector used
to calculate radial velocities (sector angle is $40^\circ$).
However the asymmetric profile along the major axis suggests that
Northern and Southern arms do not have the same radial velocity profiles.
The asymmetric tidal forces of NGC~7769 and NGC~7770 affecting on NGC~7771,
seem to be a natural cause of that.

The rotation curve of NGC~7770 is significantly skewed.
This is probably because of the strong harassing
interaction with the more massive NGC~7771, see \citet{2012MNRAS.425L..46A}.
The rotation curve of NGC~7771A is typical for a late type Sm galaxy.

By analyzing velocity fields, sizes, and shapes of
spiral arms of NGC~7771 and NGC~7769, in \citet{1997AJ....114...77N}
it has been suggested that NGC~7771 and NGC 7769, which have a 2:1
mass ratio, appear to be having a prograde-retrograde interaction,
with NGC~7769 being the retrograde one.
Our better data support this conclusion.
This conclusion is in agreement with the latest models of
galaxy collisions (\citealt{2007A&A...468...61D}) showing that during
direct collisions tidally induced spiral arms are
much longer and brighter than those during retrograde collisions.
We can conclude that galaxies NGC~7769 and NGC~7771 already have passed
the first pericenter stage, however, probably
the second encounter has not happened yet.
The first pericenter distance should have been large enough
(around few sizes of the galaxies), so that large disturbances
in rotation curves have not appeared yet.

\section{Summary}

The quartet of galaxies NGC~7769, 7770, 7771 and 7771A is a system
of interacting galaxies.
Here, we present a Fabry-Perot imaging study of the system in
H$\alpha$ line. We came to the following main conclusions:

\begin{itemize}
\item Close interaction between the component galaxies of the system has produced
morphological features that are characteristic of the interactions.
We have detected features such as tidal arms, spiral arms induced by close interaction,
bars and induced star formation.
\item From the results of our interferometric observations, we
obtained the radial velocity profiles of galaxies.
The rotation curve of NGC~7769 is weakly distorted.
The rotation curve of NGC~7771 is strongly
distorted by the tidal arms caused by
direct flyby of NGC~7769 and flyby of a smaller neighbor NGC~7770.
The rotation curve of NGC~7770 is significantly skewed because of
the interaction with much massive NGC~7771.
\item The radial velocity profiles and morphological disturbances
suggest that the NGC 7769 and NGC~7771
have passed the first pericenter stage,
however, probably the second encounter has not happened yet.
\end{itemize}

Study of such systems with methods combining photometric and visual analysis
is an effective way to clarify features of star formation
in different stages of interaction.
Ongoing and future surveys using integral field spectroscopy will
allow also to explore the spatial distribution of star formation in interacting systems.

%\bibliography{editor}  % For BibTex

% For non-BibTex:

\end{document}